\documentclass[sigconf]{acmart}
\pdfoutput=1
\usepackage{tikz}
\usetikzlibrary{shapes}

\usepackage{array}
\usepackage{arydshln}
\usepackage{csquotes}
\usepackage{makecell}
\usepackage{caption}
\usepackage{subcaption}
 
\usepackage{amsmath}
\DeclareMathOperator*{\argmax}{arg\,max}

\begin{document}






\title{Designing Fiduciary Artificial Intelligence}


\author{Sebastian Benthall}
\affiliation{
  \institution{New York University School of Law}
  \country{USA}
}
\email{spb413@nyu.edu}

\author{David Shekman}
\affiliation{
  \institution{Northwestern Pritzker School of Law}
  \country{USA}
 }
\email{david.shekman@law.northwestern.edu}


\begin{abstract}
   A fiduciary is a trusted agent that has the legal duty to act with loyalty and care towards a principal that employs them.
   When fiduciary organizations
   interact with users through a digital interface,
   or otherwise automate their operations with artificial
   intelligence, they will need to design 
   these AI systems to be compliant with their duties.
   This article synthesizes recent work in
   computer science and law to develop a procedure for designing and auditing Fiduciary AI.
   The designer of a Fiduciary AI should understand the context of
   the system, identify its principals, and assess the best interests
   of those principals.
   Then the designer must be loyal with respect
   to those interests, and careful in an contextually appropriate way.
   We connect the steps in this procedure to dimensions of Trustworthy AI, such as privacy and alignment.
   Fiduciary AI is a promising means to address the incompleteness of
   data subject's consent 
   when interacting with complex technical systems.
\end{abstract}

\maketitle
\pagestyle{plain}

\section{Overview}



Fiduciary duties are some of humanity's oldest rules \cite{frankel2011fiduciary}, 
dating back to Hammurabi's Code
and legally recognized throughout the world today.
Named for \emph{fiducia}, the Latin word for trust,
they concern the duties of an \emph{agent} hired to perform a task for a \emph{principal}.
Because the agent is often in a position of expertise or power
over the principal, the fiduciary duty legally bolsters the
loyalty and care expected of the agent beyond
the specific terms of their contract.
They apply to professional roles
such as health providers, legal and investment advisors, and
the directors of trusts and corporations.

Fiduciary duties already apply to technology companies operating in
some regulated sectors.
In the European Union, under the Data Governance Act \footnote{Regulation  2022/868 of the European Parliament and of the Council of 30 May 2022 on European data governance and amending Regulation (EU) 2018/1724 (Data Governance Act), 2022 O.J. (L 152), 1-44.
}, data intermediaries will be fiduciaries.
Legal scholars have begun exploring the 
possibility of expanding fiduciary duties to
information controllers more generally \cite{balkin2015information,khan2019skeptical,richards2020duty, balkin2020fiduciary}.
These broader applications of fiduciary principles have appeared in
recent legislative proposals.\footnote{Federally — American Data Privacy and Protection Act, H.R. 8152, 117th Cong. (2022); Data Care Act of 2021, S.919, 117th Cong. (2021). For state — H.B. 1749, 192nd Sess., (Mass. 2021); New York Consumer Privacy Act, S. 6701B, 2021-22 Sess. (NY 2021).}

Today, many companies engage their clients primarily through an automated user interface.
When companies deliver services at scale via a website, an app, or a physical object in the Internet of Things,
they delegate their operations to a computational system.
These systems are increasingly trained on data from their users
using general purpose machine learning algorithms and pretrained models for tasks like facial recognition and
natural language processing, and hence are artificial intelligence systems\footnote{For a standard definition of artificial intelligence systems, see OECD Recommendation
on AI:2019; ISO/IEC 22989:2022}.
Referring to automated clients as Artificial Intelligences (AI) can conflate the automated decisions made by at the interface with the ongoing choices made by engineers, content moderators, designers, and product managers.
However, as the client interfaces become more automated and trained using machine learning, the more referring to them as AI becomes more literally correct.

This article is a guide to designing AI systems that are operated by people or corporations that have fiduciary duties to others that are principals or beneficiaries.
We refer to an artificial intelligence that upholds fiduciary duties owed by its operator as Fiduciary AI.
\footnote{This is also what we mean when we discuss fiduciary duties for computational systems, or the application of these duties to AI.}
Fiduciary AI is one form Trustworthy AI (TAI), a way of designing intelligent systems that are lawful, ethical, and technically and socially robust.

The first thing a designer of a Fiduciary AI needs is a working understanding of what fiduciary duties are and why they have been applied.
Section \ref{sec:fd} presents the legal arguments for fiduciary duties in general.
Fiduciary duties are invoked when an agent holds power over its principals and when the principal is unable to completely specify the actions of the agent through consensual contract.
These conditions hold in many applications of AI.

To the extent possible in an area of unsettled law, Section \ref{sec:to-practice} discusses what measures should be taken by the designer of an artificial intelligence system in order for it to be compliant with these laws.
We connect aspects of Fiduciary AI to known dimensions of Trustworthy AI \cite{wing2021trustworthy, varshney2019trustworthy}, including privacy and AI Alignment. 
We outline a procedure involving a systematic series of sensitizing questions to be asked when the AI is designed or audited for compliance.
For each step in the procedure, we then identify how this aspect of Fiduciary AI has been illuminated in prior literature.

We maintain that a Fiduciary AI should be scoped to a particular context of activity (Section \ref{tai:context})
and designed with an identified set of principals in mind (Section \ref{tai:identification}). Notably, these two steps of the procedure concern legal requirements and the prioritization of beneficiaries, which are not technical questions but nevertheless logically precede questions concerning the system's technical design.

With this contextual information in hand, the system can then be designed in technical detail.
The best interest of these principals must be assessed (Section \ref{tai:measurement}), which may involve machine learnt representations of the principals' best interests.
If there are many principals, these interests must be aggregated (Section \ref{tai:aggregation}).
Fiduciaries are ultimately bound to duties of loyalty and care.
Fiduciary AI loyalty can be understood as alignment with the assessed best interests of the principals (Section \ref{tai:loyalty}).
Care refers to the adherence to other best practices of community norms guiding the design of the system, such as the determination of the systems' inductive bias (Section \ref{tai:care}).




\section{Fiduciary Principles and Regulatory Trends}
\label{sec:fd}

Fiduciary duties are, foremost, legal constructions that encapsulate and enforce expectations of trustworthiness.
This section discusses fiduciary law and is organized as follows.
First, we will discuss fiduciary duties in general as they are understood in many sectors that predate current uses of computation and AI, with specific example of duties that can inform Fiduciary AI design.
Next we will discuss proposed expansions
of fiduciary duties to new computational domains.
The application of fiduciary duties to computational systems is currently an area of unsettled law. We do not in this paper take a position on which systems ought to be legally held to fiduciary principles.
However, because legal outcomes are inherently normative and discursive, we present the broad legal debate to best anticipate the engineering and design principles that would guide a TAI system toward compliance.

We must note that fiduciary laws vary between jurisdictions.
This paper will utilize the language of Anglophone common law language to describe fiduciary relationships, principles, and actors,
and focus on U.S. and European jurisdictions.
The term ``fiduciary duties'', as it is understood in the Anglophone legal context, is difficult to translate directly across all languages.
However, the duties of loyalty and care which comprise the concept have direct corollaries in the Western European context \cite{Gelter2019OxfordEuro}. 
Moreover, while there are significant differences between common law and civil law variations of fiduciary duties, there has been a convergence of principles with respect to \emph{new} fiduciary duties aimed at actors in the data economy and, by extension, AI.
We find that the key principle of Fiduciary AI is that it must be aligned with the best interests of the beneficiaries in order to remedy an imbalance of power and necessary incompleteness in contracting.

\subsection{Justifications of fiduciary principles}

Fiduciary duties arise from a recognition that specialization is an important factor in society.
It would be nonsensical and rather inefficient for every person to gain the expertise necessary to become a doctor, lawyer, or financial advisor every time they need one \cite{frankel2011fiduciary}. 
Instead, society relies on those individuals with whom that expertise is concentrated.
However, this creates a risk for beneficiaries.
Fiduciary duty dictates that fiduciaries must act with loyalty and care to their beneficiaries and face stiff penalties if they violate these duties.
Fiduciary duties are justified in legal scholarship according to at least two lines of argument —
an argument based on the vulnerability of the principal in fiduciary relationships
and an argument about economic efficiency under conditions when transaction costs make contracting necessarily incomplete.

\subsubsection{Fiduciary relationships}
\label{sec:fid-rel}

The most common account of fiduciary duties focuses on the relationship between
the principal and the agent,
where the asymmetry of knowledge and power creates a fiduciary relationship \cite{miller2019identification}.
That asymmetry comes from what \citet{frankel2011fiduciary} calls ``entrustment''. Patients trust doctors with their bodies, clients trust financial advisors and lawyers with their money and discretion.
This entrustment of power over oneself or on one’s behalf raises the potential for harm by the trustee.
Because of the vulnerability this creates, additional restrictions beyond commercial transactions’ standard obligation of good faith have been established to protect the principals. 

Certain roles or statuses have been deemed to be fiduciaries by 
conventional wisdom (and common law),
and fiduciary principles attach to agents by virtue of membership in such status. 
\citet{miller2019identification} notes trustees, directors, agents, lawyers, and doctors as well-known examples. 
Courts may also apply to fiduciary duties in other cases wherein the facts bear
that the agent's role bears significant enough resemblance to a fiduciary, for example because
they cultivate a relationship of trust, hold themselves out as an expert in the field, 
are relied on by a person for advice, and have that person's complete trust \cite{kelly2019factbased}. 

%


\subsubsection{Contractarian justification}
\label{sec:l-and-e}

Parallel with the fiduciary relation arguments, the Law and Economics (L\&E) branch of legal scholarship has endorsed fiduciary duties on different grounds.
\citet{easterbrook1993contract} leave morality out of the equation, aligning more closely with a contractarian approach.
Contractarians view the entire 
commercial milieu, both agency relationships and the 
corporation itself, as composed of contracts \cite{jensenmeckling}.
By default, the design burden lies with the contracting parties.
\citet{easterbrook1993contract} view the courts as 
playing an active role in specification of the contract via 
interim or ex post adjudication. 
The court functions as the theoretical completing 
piece of the incomplete contract between the parties \cite{brooks2019economics}.

This is especially important in cases where there is a severe imbalance of transaction costs against the principal party.
Monitoring and specification costs, it is posited, would prevent the negotiation of contracts to completion.
But the duty of loyalty (intention) and duty of care (execution) establish, in an incomplete contract, the standards that would be arrived at were full negotiation possible.
The expertise of the agent is relied on to ``complete'' the scope of the contract according to those duties; the agent is expected to charge a premium for foregone opportunities that result from the duties.
Thus the principal can engage the transaction trusting to get what they wanted from it: the agent's expertise. 

Fiduciary duties are a way to solve the principal-agent problems of moral hazard and hidden action. Rather than disempower the agent, fiduciary duties allow the principal an \textit{ex post} review of the agent’s behavior. The duties are written as broadly as possible in order to cover the territory left open by incomplete contracts \cite{sitkoff2011economic}. 
This enables contracts that would have otherwise failed due to the large transaction costs \cite{easterbrook1993contract}.


\subsubsection{Contextuality and subsidiary duties}
\label{sec:contextuality}

In the United States, the primary fiduciary duties are the duty of loyalty and duty of care.
These are broad, open-ended standards used across many sectors.
In practice, these broad duties are interpreted by courts into
more granular \emph{subsidiary duties} that are
``field-specific elaborations'' \cite{sitkoffother}
of the primary duties.
The definitions of subsidiary duties crystallize the interpretation 
of broader duties of loyalty and care into forms that are easier 
to comply with and enforce.
Table \ref{tab:subsidiary} displays subsidiary duties for
a variety of professional statuses to which fiduciary duties apply,
as well as speculative subsidiary duties that might apply to computational systems.
In new fiduciary duties introduced by statute, subsidiary duties may be explicit legal clauses that accompany expressions of the more general duties. 
Often these subsidiary duties refer to the flow of information from the agent to or about the principal.

\begin{table*}[t]
  \centering
  \begin{tabular}{l|l|l|l}
    Context & Loyalty & Care & Both \\
    \hline
    \hline
    Trusts & \makecell[tl]{Incurring only reasonable \\ costs} & Prudent investor rule & \makecell[tl]{Giving account to\\beneficiaries$^\dagger$} \\
    & & & \\
    &  & \makecell[tl]{Duty to not commingle trust\\property} & Record-keeping$^\dagger$ \\
    \hline
    \makecell[tl]{Corporate \\ management} & \makecell[tl]{No usurpation of corporate\\opportunity}  & \makecell[tl]{Need for monitoring and\\compliance} & \makecell[tl]{Disclosure to\\ shareholders$^\dagger$} \\
    & & & \\
    & \makecell[tl]{No impairment of shareholder\\meetings} & & \\
    & & & \\
    & \makecell[tl]{Boardroom confidentiality$^\dagger$} & & \\
    \hline 
    \makecell[tl]{Investment \\ advice}& \makecell[tl]{Best execution of instructions} & Prudent investor rule & \makecell[tl]{Keeping books and\\records$^\dagger$} \\
    \hline
    \makecell[tl]{Legal \\ representation} & \makecell[tl]{No conflicts of interest} & \makecell[tl]{Safeguarding the client's\\confidences$^\dagger$} & \makecell[tl]{Communication with\\the client$^\dagger$}  \\
    \hline
    Health care & Confidentiality$^\dagger$  & & \makecell[tl]{Informed consent$^\dagger$} \\
    \hline
    \makecell[tl]{(DGA) Data \\ Intermediaries$^\ddag$} & Facilitate exercise of rights & Security and confidentiality$\dagger$ & Notice of data uses$\dagger$
    \\
     & Act in subject's best interest & Ensure interoperability & Consent management \\
    \hline
    \hline
    \makecell[tl]{Data \\ Processors$^\ddag$}  \cite{hartzog2022legislating} &  & &  \\
    & & & \\
    \makecell[tl]{Collection} & Data minimization & Proper disposal$\dagger$ & Data records$\dagger$ \\
    & & & \\
    \makecell[tl]{Personalization} & \makecell[tl]{No conflict with collective \\ best interests}  & Eliminate disparate impact & \\
    & & & \\
    \makecell[tl]{Gatekeeping} & \makecell[tl]{No exposing principals to \\ privacy harms} & Counterparty evaluation & Proper security practices$\dagger$ \\
    & & & \\
    \makecell[tl]{Influence} & Eliminate dark patterns & \makecell[tl]{} &  \\
    & & & \\
    \makecell[tl]{Mediation} & Disincentivize user harms  & Mediation algorithms oversight &  \\
    & & & \\
    & \makecell[tl]{Hierarchical loyalty for\\heterogeneous roles} & & \\
    \hline
    \makecell[tl]{AI \\ Assistants$^\ddag$} \cite{aguirre2020ai} &  No conflicts of interest
    & \makecell[tl]{Clearly indicate potential\\conflicts$\dagger$}  & \\
    & & & \\
    & Transparent objective functions & Adequate requests of user input$\dagger$ & \\
    \hline
  \end{tabular}
  \caption{Examples of subsidiary duties by legal field.
  Includes hypothetical examples of subsidiary duties for information processors. Subsidiary duties marked with a dagger ($\dagger$) concern information flows.
  Contexts marked with a double dagger ($\ddag$) are fiduciary contexts not yet determined by law
  but suggested in the literature \cite{aguirre2020ai, hartzog2022legislating}.}
  \label{tab:subsidiary}
  \vspace{-2em}
\end{table*}

\subsubsection{Example: the prudent investor rule}
\label{sec:prudent-investor-rule}

An example of a subsidiary duty is the \emph{prudent investor rule}, a subsidiary
duty of care in the context of trust law.
\footnote{Originally, the ``prudent man rule'' was a common law standard
enshrined in 1830 (\emph{Harvard College v. Amory}) that a trustee prioritize regular income over
speculative value when managing trust assets.
Later, this was updated in the American Law Institute's Restatement Third of Trusts and promulgated as the Uniform Prudent Investor Act (UPIA).
(Restatement (Third) of Trusts § 90 (2007),
61 A.L.R.7th Art. 1)}
This rule, which has since
been adopted by many U.S. states, aligns the fiduciary duties of a trust with modern portfolio theory \cite{markowitz1968portfolio}, which is a 
scientifically validated standard of asset management.
As a scientific expert standard, it generalizes the idiosyncratic interests of trustees in a way that is consistent with the purpose of the fiduciary as a trustee.
The trustee is bound by the duty of care to this formulation of the principals' interests, while the duty of loyalty requires them to put the principal's interests over their own.

\subsubsection{Example: the duty of impartiality}
\label{sec:conflicts}

One example of a subsidiary duty that arises from the duty of loyalty, which will be relevant to the design of Fiduciary AI, is the 
\emph{duty of impartiality}, which applies to trusts and corporate directors.
This duty comes into play when there may be
conflicting interests between multiple beneficiaries \cite{golddefense}, such as trustees with conflicting preferences \cite{schwarcz2009fiduciaries} or different classes of shareholders \cite{nir2020one}.
Impartiality does not mean that fiduciaries must treat each beneficiary equally, merely that their balancing of beneficiaries’ interests is not influenced by the fiduciary’s self-interest or favoritism. 
The corporate directors can change the weights they assign to the preference sets of differing shareholders according to circumstance \cite{nir2020one}.
We discuss this further in Section \ref{tai:aggregation}.

\subsection{New computational fiduciaries}

Recent scholarship, advocacy, and legislation has called for extending fiduciary duties to new contexts
associated with AI and digital services.
We briefly consider some of the arguments and controversies of these proposals here.

\subsubsection{As a remedy for incomplete contracting.}
\label{sec:incomplete-contracting}

One key motivation for extending fiduciary duties to
AI and digital services is the insufficiency of the
``notice and consent'' framework currently governing consumer privacy and data protection more broadly.
Users of a commercial digital service sign a contract with the provider that determines the terms of their interaction, including the use of their data.
It is well known that most users will not read the contracts that they digitally sign \cite{mcdonald2008cost, obar2020biggest}.
In some cases, the system's design can be such that the offered consent can be considered either uninformed (c.f. \cite{friedman2000informed, romanosky2006privacy}) or manipulated \cite{calo2013digital, susser2019technology, bongard2021definitely}.
Such a design need not be intentional.
An artificially intelligent system trained to optimize
an objective function (such as total clicks on advertisements) can \emph{learn} to present an interface to users that elicits their uninformed consent.
In short, the burden of the transaction costs to users of asserting their interests through contracting prohibitively high, and AI operating with conflicted interests can exploit this.
Corroborating this view, \citet{hadfield2019incomplete} have drawn the connection between AI Alignment and the legal challenges around incomplete contracting, and conclude that AI Alignment will depend on a larger legal framework in which a community's normative structure is imputed into the terms of the relationship between the principal and the agent.
As discussed in Section \ref{sec:l-and-e}, fiduciary duties are an available legal tool for establishing
the alignment of an agent to a principal when contracting is incomplete.

\subsubsection{Example: AI assistants}

\citet{aguirre2020ai} consider specifically the case of AI assistants like Alexa, Siri, Google, and Cortana which serve as interfaces between consumers and the web.
While these assistants may appear to act in the interests of their users, they may be designed with embedded conflicts of interest.
Duties of loyalty can correct this.

\subsubsection{Example: E.U. Data Governance Act (DGA)}

The European Union recognizes a fundamental right to data protection \cite{schwartz2017transatlantic}.
In 2016, the EU passed a landmark data protection law, the General Data Protection Regulation (GDPR), which received global attention due to its strong extraterritorially enforced sanctions \cite{streinz2021evolution}.
Under the GDPR, consent is a legal basis for lawful processing of personal data that is often invoked in commercial applications.\footnote{Article 6 Regulation (EU) 2016/679 of the European Parliament and of the Council of 27 April 2016 on the protection of natural persons with regard to the processing of personal data and on the free movement of such data, and repealing Directive 95/46/EC (General Data Protection Regulation), art. 6, OJ 2016 L 119/1.} 
This includes data collected about internet browsing behavior through cookies, which is a key data source underpinning targeted advertising. Consent Management Platforms (CMPs) are software solutions that collect, 
store, and monitor users' consent to uses of this personal data \cite{santos2021consent}.
Use of these CMPs has expanded rapidly since the passing of GDPR \cite{hils2020measuring}. However, these CMPs have not successfully guaranteed data protection rights as the GDPR has intended.
\footnote{The Transparency and Consent Framework (TCF) \cite{iab2020tcf}, a standardized CMP developed by IAB Europe (an association for the digital marketing industry), was found to violate GDPR by Belgium's Data Protection Authority on several grounds. (Belgian Data Protection Authority, Decision on the Merits 21/2022 of 2 February 2022, Complaint Relating to Transparency \& Consent Framework (IAB Europe), DOS-2019-01377, \url{https://www.autoriteprotectiondonnees.be/publications/decision-quant-au-fond-n-21-2022-english.pdf})
Others have frustrated privacy scholars and users alike by employing 
dark patterns and choice architecture to push users toward invalid or uninformed consent \cite{nouwens2020dark}.
Indeed, researchers have found that in widely deployed CMPs,
data is collected, processed and shared even when users have not consented to it \cite{lieu2022opted}.}

European lawmakers have passed the Data Governance Act (DGA), which addresses the uncertain legal status of CMPs by identifying them as a type of ``data intermediary''. \footnote{\emph{See} Recital 30, Regulation  2022/868 of the European Parliament and of the Council of 30 May 2022 on European data governance and amending Regulation (EU) 2018/1724 (Data Governance Act), 2022 O.J. (L 152), 11.} 
According to the DGA, ``data intermediation services providers seek to enhance the agency of data subjects, and in particular individuals' control over data relating to them''.\footnote{Recital 30, Regulation  2022/868 of the European Parliament and of the Council of 30 May 2022 on European data governance and amending Regulation (EU) 2018/1724 (Data Governance Act), 2022 O.J. (L 152), 11.
}
The Act uses the language of fiduciary duties and ``act[ing] in the best interests of the data subjects''.\footnote{``Data intermediation services providers that intermediate the exchange of data between individuals as data subjects and legal persons as data users should, in addition, bear fiduciary duty towards the individuals, to ensure that they act in the best interest of the data subjects.'' Recital 33, Regulation  2022/868 of the European Parliament and of the Council of 30 May 2022 on European data governance and amending Regulation (EU) 2018/1724 (Data Governance Act), art. 12(m), 2022 O.J. (L 152), 12-13.
} 

Under this law, data intermediation is its own fiduciary context.
The DGA restricts data intermediaries to using collected data for only the purposes of its intermediation services.
It further prohibits the pricing and other commercial terms of the service from being dependent on the customer's degree of use of any other service. 
In short, data intermediaries under the DGA are fiduciaries that assist individuals in exercising their data rights for their best interests.

\subsubsection{Information fiduciaries and their discontents}
\label{sec:information-fiduciaries}

While fiduciary principles are already relevant to computational systems
operated in sectors regulated by existing fiduciary law,
and to data intermediaries by the DGA,
recent work has proposed the expansion of fiduciary principles
to data processors more general, creating so-called ``information fiduciaries'' \cite{balkin2015information} with ``data loyalty'' \cite{hartzog2022legislating}.
Several bills have been proposed both at the federal and state level aiming to apply duties of loyalty and care, as well as subsidiary duties, to platforms.\footnote{Federally — American Data Privacy and Protection Act, H.R. 8152, 117th Cong. (2022); Data Care Act of 2021, S.919, 117th Cong. (2021). For state — H.B. 1749, 192nd Sess., (Mass. 2021); New York Consumer Privacy Act, S. 6701B, 2021-22 Sess. (NY 2021).}
\footnote{This trend is not limited solely to the US. In India, the 2021 Data Protection Bill, currently under discussion by the parliament \cite{mandhani_2021}, explicitly creates data fiduciaries and data principals. Though the language of data fiduciaries is used in this bill, the functional implications appear to align closer with that of the ``data controller'' under the GDPR regime. \cite{trilegal}}

Broad information fiduciary laws are controversial.
They would upset how many well-established multi-sided platform companies \cite{cusumano2019business}, which necessarily navigate conflicting interests of multiple stakeholders, do business -- and hence how they would use AI, such as recommendation systems.
This raises legal questions about how an information fiduciary would manage conflict duties to principals in different roles, such as its users, advertising partners and shareholders \cite{khan2019skeptical, grimmelmann2019when}.
However, corporate fiduciaries are still bound by the measures of the law \cite{tuchdefense} .
Committing an illegal act would be detrimental to the financial value of the company (either through impositions of fines or other penalties, reduced revenues, or loss of brand value/goodwill)
and thus the shareholders' interests.
If fiduciary duties to users of platforms were enacted into law,
then that would be another legal constraint within which 
the platforms' corporate officers would need to act.
We consider the prioritization of multiple beneficiary roles in Sections \ref{tai:identification} and \ref{tai:loyalty}

\section{Engineering Fiduciary AI Systems}
\label{sec:to-practice}

For the reasons discussed in Section \ref{sec:fd},
many technology businesses may find that they have fiduciary duties.
Given that businesses increasingly rely on automation and AI for their operations,
and that they can be accountable to fiduciary duties for these operations,
these businesses should be aware of how their automated systems
can be designed \emph{ex ante} to avoid violation of their legal
obligations.
To this end, some technical interpretation
is helpful for identifying what constitutes compliance
with the law and to what extent compliance can be guaranteed
by an auditor.
\footnote{Of course, it may be that no \emph{ex ante} design will be sufficient for compliance with fiduciary duties, given the possible \emph{ex post} actions of businesses and the variability of judicial
interpretation.
Some legal scholars are openly skeptical about the possibility of engineering compliance \cite{cohen2016regulatory, waldman2019privacy, waldman2021industry}.
However, we will connect fiduciary duties to Trustworthy AI principles and techniques as a pragmatic exercise.
We acknowledge that there are both limits to \cite{selbst2019fairness} and value in \cite{abebe2020roles} the use of computational abstraction to clarify normative guidelines.
But for AI systems in particular, designers \textit{must} make ethical choices.
For example, \citet{baum2020social} argues that there are three sets of decisions in a social AI system that require the imposition of the designer's values: standing, measurement, and aggregation.
We find that analogous questions arise when designing a fiduciary AI. We have replaced ``standing'' with ``identification'', and ``measurement'' with ``assessment''. Standing, in Baum’s context, is ``to have one’s ethics included in a social choice process used to determine the ethics of AI'' \citet{baum2020social}. However, this differs from the use of the term ``standing'' in law to refer to whether one has the ability to bring suit.}

The remainder of this paper assumes that a technologist is
interested in implementing a fiduciary service or artificial agent.
What do fiduciary duties mean for their technical practice and design?
We contribute a procedure for a system architect to consult when attempting to build a new system that is compliant-by-design with fiduciary duties, or when auditing an existing system for compliance.
It is formulated as a rubric of questions.
(See Table \ref{tab:rubric}.)
The questions are presented in a logical sequence such that the answer to early questions, such as ``What is the context of the system?'' and ``Who are the principals?'' inform the answers to later questions, such as ``Is the system aligned with the best interests of the principal?''.

While many of the designers decisions are quite technical, we find that subsidiary fiduciary duties may directly inform the design at many stages.
For example, a reasonable person standard from a subsidiary fiduciary duty can inform the answer to the question, "What are the best interests of the principals?", which would otherwise be derived from data.
In this way, subsidiary duties may guide \emph{ex ante}
expectations of behavior and guide designs.
However, fiduciaries will also be held to the 
more abstract, primary fiduciary duties of loyalty and care.
These latter duties allow for flexibility by the courts and regulators in \emph{ex post} analysis.

Further sections of this paper address facets of our rubric for Fiduciary AI by outlining corresponding techniques or frameworks from computer science.
Naturally, in an area of unsettled law, using such a rubric will be necessary at best, and never sufficient, for compliance with regulations.\footnote{In the future, such sufficient conditions could be defined by law via a ``safe harbor'' provision, which is essentially a well-understood floor of expectations above which a covered entity will be considered to be compliant.}
And in many cases, the TAI dimensions implicated by fiduciary duties are active research areas with many open questions.
We see Fiduciary AI as an emerging area of both law and engineering.
What follows is a survey of the state of the art at the intersection of these fields.

\begin{table*}[t]
  \centering
  \renewcommand{\arraystretch}{1.3}
  \begin{tabular}{lll}
    1. Context & What social context or sector will the system will operate in? & Section \ref{tai:context} \\
    & What are the purposes of that context? & \\
    & What roles are defined in that context? &\\
    & What norms or rules does that context imply? & \\
    \hline
    2. Identification & Who are the principals? & Section \ref{tai:identification} \\
    & Is there more than one category of principals? & \\
    & Do the categories have different levels of priority? & \\
    \hline
    3. Assessment & How are the interests of the principals being assessed? & Section \ref{tai:measurement}\\ 
     & Are they being empirically measured? With what data? & Section \ref{sec:loyalty-unknown} \\
     & Are future rewards discounted? By how much? & Section \ref{sec:temporal}\\
     & Is there are reasonable person standard of best interest? & Section \ref{sec:assess-legal} \\
     \hline
    4. Aggregation & If there is more than one principal, how are their interests aggregated? & Section \ref{tai:aggregation} \\ & Whose interests dominate in a conflict? &  \\
     & Are there duties about aggregation, e.g. impartiality? & Section \ref{sec:agg-legal}\\
     \hline
    5. Loyalty & Is the system aligned with the principals' interests? & Section \ref{tai:loyalty} \\ 
     & Have subsidiary duties of loyalty been considered? & Section \ref{sec:loyalty-subsidiary} \\
     \hline
    6. Care & What is the context-appropriate level of prudence? &  Section \ref{tai:care} \\ 
    & Is the inductive bias of the system up to contextual standards? & Section \ref{sec:inductive}\\
  \end{tabular}
  \caption{Steps for designing or auditing a Fiduciary AI for compliance, with corresponding sections of this paper where they are discussed.}
  \label{tab:rubric}
  \vspace{-2em}
\end{table*}

We see Fiduciary AI practice as involving a novel combination of Contextual Integrity \cite{nissenbaum2020privacy}, a theory of ethical computing design that focuses on contextualized social norms, and the problem of AI Alignment.
The AI alignment problem, or how to design an AI that acts
in a way that is aligned with the intentions of principal users,
has been most widely discussed in the growing field of Artificial General Intelligence (AGI) research \cite{dewey2011learning, hadfield2016cooperative, everitt2021reward}.
Fiduciary duties are one form of legally binding, contextualized normative structure that can address the necessary incompleteness in
the relationship between AI and its users \cite{hadfield2019incomplete}.
Thus, Fiduciary AI design can be a precursor to a better understanding of Aligned AI \cite{aguirre2020ai}.

\section{Context}
\label{tai:context}

In the law, there are no absolute fiduciaries who are required under all circumstances to serve the interests of a principal.
Rather, individuals are fiduciaries to others by virtue of their respective roles in a legally recognized context.
A first design principle for Fiduciary AI is that the context within which the AI is acting as a fiduciary must be established and understood.

The study of how context should inform computational system design has deep roots in ubiquitous computing research \cite{ackerman2001privacy, dey2001conceptual, dourish2004we}.
We recommend that Fiduciary AI draw on
Nissenbaum's Contextual Integrity (CI) \cite{nissenbaum2020privacy}, a well-developed theory of social contexts that is used in the design and evaluation of technical systems\footnote{Contextual Integrity is mainly about norms of information flow, and defines privacy as appropriate information flow. Each norm is parameterized in terms of actors (sender, receiver, and data subject); attributes (meanings of information) and transmission principles (for instance confidentiality, reciprocity, or ``with a warrant''). The analyst is encouraged to evaluate the introduction of a new technology with a heuristic procedure that identifies how any new information flows depart from the norms embedded in the context of interest. One of the steps in the procedure is to evaluate any changing information flows in terms of the purposes of the context as it functions in society more broadly. Ultimately, the norms are animated by the specific purpose of that context, as well as societal values and individual goals.}.
It begins from the position that a pluralistic society \cite{walzer2008spheres} can be understood to consist of multiple social contexts or spheres, each with its own internal system of roles, rules, and meanings.
CI offers a rubric for understanding contexts in terms of:
\begin{itemize}
    \item \emph{Purposes}. The purpose of the social context in society, towards which its laws and norms are instrumental.
    \item \emph{Roles}. The names and definitions of roles that agents can have in that context.
    \item \emph{Norms}. Social or legal rules prescribed within the context about how agents of the identified roles should interact.
\end{itemize}

Translating the CI framework into technical requirements is an ongoing area of computer science research \cite{benthall2017contextual}
We do not recommend any particular implementation of CI, but rather recommend using it to frame the answer to a specific question: in what context is an AI expected to be performing fiduciary duties?
This understanding of context in terms of roles, purposes, and norms will be used to answer other questions related to identification (Section \ref{tai:identification}) and aggregation (Section \ref{tai:aggregation}).
Furthermore, CI is a fitting framework for evaluating the context of fiduciary duties because some subsidiary duties, like duties of disclosure and confidentiality,
are explicitly about information flow.
We develop this case as an example of alignment in Section 
\ref{sec:loyalty-subsidiary}.

\section{Identification}
\label{tai:identification}

Designing Fiduciary AI requires an explicit understanding of who a system's principals are.
The question ``who are the beneficiaries of this system?'' has long been one of the first questions to ask when designing any social system \cite{ulrich1987critical}.
In most cases today, AI is designed mainly to benefit those that deploy the system, using mechanism design techniques to steer the behavior of users towards those goals \cite{viljoen2021design}.
Indeed, some of the literature on AI alignment assumes that the goal of AI safety is to ensure that the AI system is aligned with the interests of its operator \cite{amodei2016concrete}, while others defer the question of the ``preference payload'' to other research \cite{leike2018scalable}.

Fiduciary AI can be designed with other categories of principals in mind, and identifying these principals is
essential for compliance with fiduciary duties.
This choice cannot be founded on purely technical considerations, rather, it must be inferred from the previously identified legal or social context (see Section \ref{tai:context}).
For example, many companies that build AI products
operate under Delaware corporate law and their
corporate directors have only one pressing set of
fiduciary obligations: to their shareholders.
But a business that is based in Delaware but also is operating as a data intermediary in the European Union would need to prioritize its duties to its data subjects because legal penalties for violating those duties would be, in turn, bad for the shareholders (see Section \ref{sec:information-fiduciaries}).

To illustrate the significance of the fiduciary relationship, we consider a typical online system that interacts with human users in a commercial context.
In the absence of a fiduciary relationship, the human user's engagement with the system will be regulated by a contractual agreement to which the user has legally consented. (See Section \ref{sec:incomplete-contracting}.)
We note that \emph{for those users identified as principals of a fiduciary system}, consent is \emph{not sufficient} for the fulfillment of the fiduciary duty.
Rather, other requirements such as loyalty (see Section \ref{tai:loyalty}), must govern the interface design to ensure as much as possible that consent, when given, is in the interests of the principal.

In a similar vein, consider a user who has consented to a system's broad terms and conditions and now operates the system through a user interface.
Normally, a system would be expected to perform whatever actions the user chooses for it to do.
However, for a complex system, the instructions given will never be a complete description of the operations to be performed,
and the choice of instructions may be manipulated by a system's design.
Indeed, there is a division in law between two models for the interpretation
of the duty of loyalty — the \emph{obedience} model, 
wherein the fiduciary must follow directions given by the beneficiary,
and the \emph{best-interests model}, wherein the fiduciary acts with more discretion but in service to the beneficiary.
The latter best-interests model is favored by \citet{richards2020duty} for privacy and information platform cases, as it is precisely the complexity of the technical systems and inability of consumers to grasp their mechanics which makes contracts (as well as instructions) incomplete or impossible to monitor.
This requires judicious design choices on the part of the data controller, e.g., the choice of an email provider to maintain deleted emails on record for a short period of time in case the user ordered their deletion by accident.

In summary, while most digital or AI systems will be operated by users on the basis of consent and obedience, Fiduciary AI imposes a higher standard on the system's design with respect to its \emph{principal} users.
For example, a medical AI might be designed to serve the best interests of its medical patients (the principals), but have a different interface used by system operators (not principals) that is more strictly obedient.
For the remainder of this article, we will assume that the principals have been identified and the Fiduciary AI designer must now determine how to fulfill the system's duties to those principals.

\section{Assessment}
\label{tai:measurement}

Once the principals of a fiduciary AI system have been identified, system designers should assess the salient best interests of those principals.
This is not a trivial problem by any means: how many people can judge what is in another person's best interest, or even their own?
Luckily, this problem is constrained;
the system designer must only assess the best interests of the principal in their role within the fiduciary context.

One approach to assessing these best interests is to learn an objective function from data provided by the population of principals.
The data may be observations of behavior, statements of preferences, or direct measurements of user well-being.
We will discuss these data sources and their associated challenges.

Beyond empirical assessments of best interest, the law also furnishes some fiduciary contexts with \emph{reasonable person standards}, which are legal codifications of the interests of a normal or general person in a given role.
In Fiduciary AI cases, a combination of empirical data collection, machine learning, and reasonable person standards may be used to determine principal best interests.\footnote{While some researchers have found improved results using observed human behavior or feedback to train a policy function directly instead of training a reward function \cite{knox2010combining, griffith2013policy}, we do not recommend this as a way of achieving Fiduciary AI, because Fiduciary AI is specifically concerned with the best interests of the principals. We also do not recommend, for this application, inferring reward functions from observable states of the world \cite{shah2019preferences}, as Fiduciary AI has been proposed as a solution to problems with the status quo due to, for example, market failure.}
Once the information about each principal's best interests are identified,
they can be aggregated if appropriate (Section \ref{tai:aggregation}), and then rendered into technical design (Section \ref{tai:loyalty}), perhaps as an objective function or reward model \cite{leike2018scalable}.

\subsection{Learning reward models}
\label{sec:loyalty-unknown}

One strategy for assessing the best interests of principals is to attempt
to learn a proxy model of those interests from observational data.
This approach has been best explored in the context of reinforcement learning.
While we do not limit the scope of this work to reinforcement learning, we nevertheless point to reward modeling \cite{leike2018scalable} or reward design \cite{hadfield2017inverse} as key elements of Fiduciary AI.

Consider the Bellman equation form of the intertemporal decision problem:
\begin{equation}
\label{eq:V}
V(s) = \max_a r(s, a) + \beta \mathbb{E} \left[ V(s') | s, a \right]
\end{equation}
Where $V$ is the value function for a state $s$, given an optimal action $a$.
An implied probability distribution $P(s' | s, a)$ governs the state transitions.
The value function is defined recursively as the sum of immediate rewards $r(s, a)$ and the expected value of future rewards $\mathbb{E} \left[ V(s') | s, a \right]$, discounted by a factor of $0 < \beta < 1$.

We define $Q$ as the state-action value function:
\begin{equation}
\label{eq:V}
Q(s, a) = r(s, a) + \beta \mathbb{E} \left[ V(s') | s, a \right]
\end{equation}
An agent's policy $\pi$ maps from states to actions. 
In the reinforcement learning context, the policy is trained to optimize the action-value function.
\begin{equation}
    \pi^*(s) = \argmax_a Q_{\pi}(s, a)
\end{equation}

Researchers working in AI Alignment have worked on techniques for learning a proxy reward function $r$ that reflects the true objectives of a system's principals.
This function can then be used in training the AI system's policy through reinforcement learning.

Reward learning comes with many challenges.
In all forms of reward design, the proxy reward function
will, at best, be based on observations of the true reward
function, leaving the possibility for misalignment due to 
incompleteness \cite{zhuang2020consequences}.
It is likely that the best approach will combine several different, complementary forms of human feedback, perhaps guided by a common framework \cite{jeon2020reward}.
On-line reward learning, wherein the reward function is continuously trained on data learned through its operation, opens the learning process up to manipulation by the system during training \cite{fickinger2020multi, armstrong2020pitfalls}, so it is best to be avoided.
Furthermore, each data source has associated threats to its validity as a measure of best interests.
And in all cases, the inference of the proxy function will have inductive bias \cite{xu2019learning} (see Section \ref{sec:inductive}).

\subsubsection{Observations of behavior}

Learning the objective function of an agent based on observations of their behavior
is an enduring research problem.
One early formulation of this problem is the \emph{inverse reinforcement learning} (IRL) paradigm \cite{ng2000algorithms, abbeel2004apprenticeship, ziebart2008maximum} .
In the simplest formulations of inverse reinforcement learning, a subject's behavior is observed and this is interpreted as a policy $\pi^*$ that corresponds to their value function $V$. In principle, this allows the analyst to derive the subject's reward function $r$.

IRL faces a number of challenges.
\citet{ng2000algorithms}'s original analysis showed that IRL suffers from the problem of \emph{degeneracy} -- that there are many reward functions for which an observed policy is optimal. 
\citet{gleave2020quantifying} have shown that reward functions can be grouped into equivalence classes based on their effects on policy, which helps quantify distances in the space of reward functions.
But in any case, the outcome of IRL will depend on inductive bias (Section \ref{sec:inductive}) as well as assumptions about temporal discounting (see Section \ref{sec:temporal}).

Another set of challenges concerns how well the behavior of users reflects their
best interests.
Indeed, one might want a fiduciary service precisely when one is
\textit{unable} to act (or represent) in one's best interest.
And as a matter of general fact, people do not always act in their best interest due to cognitive biases \cite{kahneman2006developments}.
While there have been some efforts to adapt IRL to a 
``two system'' cognitive architecture \cite{peysakhovich2019reinforcement},
it has been shown that, in general, reward functions in an IRL context are unidentifiable if the principal's behavior is irrational \cite{armstrong2018occam}.

A promising direction for IRL is Cooperative Inverse Reinforcement Learning,
wherein the human trainer is an expert who deliberately teaches the system its objectives \cite{hadfield2016cooperative}.
In other words, the system is not trained on behavioral data from ``in the wild''.
Behavioral data, and especially expert training data, may be be most useful in practice when combined with other sources.


\subsubsection{Preference judgments}
\label{ref:reward-learning}

Another way to gather data about principal's objectives is to solicit their
explicit preference judgments.
\citet{noothigattu2018voting} design a system for learning ``trolley problem'' ethics by
collecting data from human users about their preferences over of different outcomes.
In the context of deep reinforcement learning, objective functions have been trained from
human judgements about the value of fragments of the system's behavioral trajectories \cite{christiano2017deep}.
These fragmentary judgments are then aggregated and recomposed into a complete objective function.
Preference judgments can be fruitfully complemented with other sources of information such as expert demonstrations \cite{ibarz2018reward}.
Researchers have experimented with grounding natural language commands into the reward function of agents in an IRL context \cite{squire2015grounding, bahdanau2018learning, fu2019language}.

A challenge associated with using this data source is that human preferences do not have the normative properties of associated with mathematical decision-theory, such as transitivity and insensitivity to alternatives \cite{tversky1989rational}.
They are also unstable; well-known cognitive science results have demonstrated that
human preference judgments are susceptible to 
framing effects, especially framing of a decision as a chance of gain versus a risk of loss \cite{tversky1985framing}.
Social scientists who survey the preferences of broad populations have learned that for studies to be effective, subjects need to have a positive reason for valuing the accuracy of the result \cite{zawojska2017re}.
In a rare treatment of these threats to validity in the AI training setting, \citet{thomaz2008teachable} find that outcomes are sensitive to how human trainers understand their relationship with the learning robots.
More research is needed to understand the human side of learning reward functions from human preferences.

\subsubsection{Measuring well-being}

Another way to discover principal preferences is to measure their well-being
directly.
Behavioral economists such as \citet{kahneman2006developments}
have addressed the problem that actions do not reveal true preferences
and have developed alternative, independent methods of measuring well-being,
such as survey techniques.
Well-being measurements like these have been employed in psychology and human-computer interaction (HCI) studies of the users of social media platforms, 
and used to assess user interface designs
\cite{kross2013facebook, panger2018people, burr2020ethics}.
IEEE \cite{ieee2020ieee} has since published 
scientifically valid well-being metrics and a Well-Being Impact Assessment
(WIA) process. 

Well-being metrics are sometimes gathered to inform public policy \cite{graham2018well}.
They can combine both objective elements (such as health or employment) \cite{linton2016review}
and subjective elements (such as frequency and duration of positive feelings) \cite{diener2009happiness}.
Well-being measurement is therefore largely separate from the fiduciary context
and may be a noisy signal with respect to the contextualized objectives of the principal.
Nevertheless, they may be a worthwhile complement to behavioral and preferential data.

\subsection{Temporal discounting and time inconsistency of preferences}
\label{sec:temporal}

Temporal discounting presents further complications and opportunities in the assessment of principal interests.
In inverse reinforcement learning, the discount rate $\beta$ is a free parameter whose value must be either assumed or fit when deriving a human model's reward function from their behavior.\footnote{Mathematically, the discount factor $\beta < 1$ is necessary
for the sum of the infinite series of expected future rewards
to converge on a finite value.}
\citet{rothkopf2011preference} explore techniques for inferring discount factors in IRL through a Bayesian process.

However, time inconsistency is yet another way in which 
principals may be irrational,
frustrating attempts to learn their ``true'' objective function.
Empirically, people's discounting of future utility takes
the form of hyperbolic discounting, as opposed to
exponential discounting
\cite{benzion1989discount, green2004discounting, chabris2008individual}.\footnote{Hyperbolic discounting means that delays that are closer in time feel more costly than the same delays further away in time, such that, for example, one prefers to pay more to prevent a delay from today to tomorrow than they would to prevent a one-day delay a year from now (from 365 days from today to 366 days from today).
Hyperbolic discounting and closely related ideas of time inconsistency of preferences and present bias \cite{laibson1997golden, o2001choice} may help explain the ``privacy paradox,'' wherein a concern for future privacy risks can be easily subsumed in the present by a desire for instant gratification \cite{acquisti2003losses}.
We do not use the phrase ``privacy paradox'' uncritically. A robust discussion of the privacy paradox is beyond the scope of this paper. C.f. \cite{solove2021myth}}
This means that even if a discount factor is fit to human behavior or preference judgements, it may an artifact of the method than a representation of the human subject's psychologically held values.
Correcting these inconsistencies may be a positive role for a fiduciary to play as an advisor.
\citet{puaschunder2021data} argues that a data fiduciary may be duty-bound 
to correct these inconsistencies on behalf
of their principals and be more rational, or even more patient, than they are.\footnote{The financial context suggests how patience might be a virtue of fiduciary advice.
Differences in discount factor are one
macroeconomic explanation for the varied distribution
of wealth in society \cite{krusell1998income, carroll2017distribution}. 
An impatient consumer is more likely to consume more today
and save less for the future.
One beneficial role of an investment advisor fiduciary
is to nudge their client to consider the long-term implications
of their decisions, in effect advising the principal on what they
should do if they discount their future less.
A Fiduciary AI might learn a principal's reward function $R$ assuming one level of patience, and then provide recommendations to them based on the same $R$, but a greater patience $\beta$.}

\subsection{Legal constraints}
\label{sec:assess-legal}

Sometimes a legal rule can define a context-specific
sense of 'best interest' that simplifies assessment a great deal.
For example,  the prudent investor rule (see Section \ref{sec:prudent-investor-rule}) 
established that for the management of trusts, the principal's
interests are best served by investing according to modern portfolio theory.
This simplifies the fiduciary's role to one of maximizing risk adjusted returns on investment; they do not need to take into account non-monetary measures of well-being.
An information fiduciary statute could likewise construct a standard of user interests based on a combination of a strictly defined context and scientific results.

While such standards simplify the assessment process, they may allow system designers to ignore important differences in user preferences, wheres privacy preferences have been shown to be quite heterogeneous \cite{urban2014privacy, egelman2015myth, marwick2018privacy}. (An alternative approach would be to assess preferences individually, and then simplify the fiduciary's goal function through aggregation, as discussed in Section \ref{sec:agg-legal}).

Translating these legal constraints into a technical specification is a challenge in its own right.
In some cases, a translation into a formal specification language such as linear temporal logic is possible \cite{datta2011understanding}.
If the rule is better operationalized as a component of or constraint on an objective function, it may need to be trained based on expert-provided data.
For example, \citet{noothigattu2019teaching} have used IRL to learn ethical constraints for a system that is otherwise trained with reinforcement learning.

\section{Aggregation}
\label{tai:aggregation}

As discussed in Section \ref{sec:conflicts}, one of the conceptual
challenges raised by computational fiduciaries
is how a system can be loyal to the
interests of multiple, potentially conflicting principals.
A natural strategy is to aggregate the many objective functions assessed for each principle into a single
objective function \cite{noothigattu2018voting}.
However, this strategy invites criticism from
the field of social choice and voting theory, which has long raised difficulties with preference aggregation.
We survey these critiques and the proposed remedies for social choice in AI, which involved partially ordered objective functions.

The available legal logic for managing conflicting fiduciaries \cite{schwarcz2009fiduciaries, golddefense} provides additional constraints on the aggregation function that may simplify the problem.
We consider subsidiary duties of impartiality and the prudent investor rule as examples of how law can guide aggregation.

\subsection{Impossibility theorems}

Social choice, the field that theorizes how to combine many individual preferences into a collective decision, has revealed many theorems that show that it is impossible for such mechanisms to meet all of several theoretically desirable criteria. A prominent example is the Gibbard-Stratherthwaite  theorem\cite{gibbard1973manipulation}, which states that for any ordinal voting system with one winner, the rule either:
\begin{itemize}
    \item is dictatorial, meaning that there is one voter who can choose the winner, or
    \item limits the outcomes to only two possibilities, or
    \item is susceptible to manipulation through insincere ballots.
\end{itemize}
Manipulation becomes a problem when the assessment of best interests is based on the principals' own voluntary actions or expressions.
These may be untrue or misleading \cite{fickinger2020multi}.
Putting aside the question of incentive compatible mechanism design,
we can note that manipulation may be avoided by decoupling the assessment process from aggregation.
In other words, this is a reason to avoid doing best interest assessment via on-line learning \cite{armstrong2020pitfalls}.
More broadly, this and other impossibility theories make the selection
of an aggregation rule a thorny problem.
\citet{baum2020social} argues that such aggregation rules are necessarily moral choices of the designer of a social AI.
Aggregating the best interests of multiple principals
of a Fiduciary AI is a structurally similar task.

Other solutions to voting paradoxes involve a relaxation of requirement that preferences or objectives be given as a total ordering.
\citet{prasad2018social} proposes that for the highest hierarchical levels of decision-making in AI design, it is best to use an approval voting mechanism, which gives each voter a binary choice of consent for every option, rather than an ordinal preference. This form of voting is less prone to manipulation.
It takes as its goal ``consent maximization'', as opposed to ``utility maximization''.
This is not to be confused with the initial consent to use the fiduciary service, but rather is meant to be a heuristic for aggregating the interests of those already identified as principals.
\citet{eckersley2018impossibility} addresses the impossibility theorems of  \citet{arrhenius2000impossibility} and \cite{parfit1984reasons} in population ethics which show that there is no way to develop a totally ordered objective function over outcomes for a population without violating human ethical intuitions in one of several ways.
This problem can be avoided if the aggregate objective function is a partially ordered, as opposed to totally ordered.
Indeed, perhaps the most straightforward way to aggregate individual objective functions is using multi-objective optimization techniques that reveal Pareto efficient indifference curves over individual welfare.
\cite{marler2004survey, gunantara2018review}

In sum, the design of the aggregation function is an important choice with ethical implications.
The relaxation of total ordering conditions eases some of the design burden.
For the remainder of this article, we assume that objective and utility functions may be partially ordered.


\subsection{Legal constraints}
\label{sec:agg-legal}

The problem of conflicting principal interests has arisen
for non-computational fiduciaries, and in some cases it has
been settled by further legal rules.
These tend to be context-specific subsidiary duties (see Section \ref{sec:contextuality}).
We consider two examples here which may be guides to how subsidiary duties could inform Fiduciary AI in the future.

In some contexts where there are potentially conflicting principals,
the duty of loyalty can include a subsidiary \emph{duty of impartiality}.
As discussed in Section \ref{sec:conflicts}, impartiality requires
that the system not be unduly influenced by either self-interest (of the agent) or favoritism (towards one of the principals). 
This constrains the agent, but the duty of impartiality also
gives the agent the flexibility to act with discretion within those constraints.
Just as corporate directors can make decisions that unequally impact different classes of shareholders \cite{nir2020one},
an information fiduciary could make decisions that unequally impact different classes of users without violating their duties.
As a further guide to how to balance principal interests, the agent should look to the specific purpose for which the relationship with the beneficiary was established.\footnote{This resolution to the multi-principal system design runs parallel to the way CI theorizes the formation of information norms: as best fit to a balance of the role-based ends of actors and the purpose of the social context.}
So in these cases, the agent's aggregation problem is constrained in some ways, but relaxed in others.\footnote{The Trustworthy AI dimensions of fairness and privacy are both relevant to aggregation in ways that are not currently considered in fiduciary law.
In privacy scholarship, there is a growing recognition of group privacy \cite{taylor2016group, mittelstadt2017individual}, when a group of people have a collective stake in their data, perhaps because of horizontal data effects \cite{viljoen2021relational}, meaning when the data collected from one subject has implications for the interest of other subjects. There is an opportunity when designing the aggregation function to acknowledge these externalities or relational concerns and balance the weighting of preferences accordingly.
Particularly well-studied in the Trustworthy AI context are fairness criteria. These fairness metrics are primarily aimed at preventing group-based discrimination in machine learning systems. However, a Fiduciary AI designer concerned with compliance with nondiscriminatory regulations or social expectations could perhaps adapt these fairness criteria to the aggregation function as well. We pose this as a problem as one for future work.}

Other legal constraints can more directly specify the agent's aggregation function.
Consider again the prudent investor rule (see Section \ref{sec:prudent-investor-rule}), which defines principal's
interests as those identified by modern portfolio theory.
This ruling irons over what might otherwise be wrinkles due to idiosyncratic principal preferences, because those preferences may not be prudent.
So, for example, if one principal wanted their trustee to divest from certain businesses for political reasons, it would not necessarily be the trustee's duty to do so.
An information fiduciary statute could likewise construct a standard of user interests based on a combination of a strict definition of context and scientific results.

\section{Loyalty and Alignment}
\label{tai:loyalty}

Loyalty is one of the two pillars of fiduciary duty.
Following \citet{richards2020duty}, we focus on a ``best interests'' interpretation of loyalty for Fiduciary AI because instructions given to an AI will be incomplete, leaving open questions of its behavior \cite{hadfield2019incomplete}.
We now assume that through the proceeding steps of the Fiduciary AI procedure, the designer has in hand an objective function (perhaps only partially ordered) that represents the assessed and aggregated best interests of the principals.
In this section, we discuss guarantees of loyalty in more depth.
In particular, we note the legal idea of the ``loyalty two-step'':
the combination of a general duty of alignment, and contextually specific subsidiary rules with clearer requirements \cite{hartzog2022legislating}.
The general duty of loyalty bears some resemblance to the broader issue of AI Alignment, and can be characterized by optimization of a proxy objective function.
Subsidiary duties of loyalty can provide firmer constraints on system behavior, and address problems that are difficult to solve using machine learning.


\subsection{Best interest and value alignment}

The first step of the ``loyalty two-step'' is a ``no-conflict'' rule in the system design \cite{hartzog2022legislating}.
This means that the system must not be designed to conflict with the best interests of the principals.
The designer must use their representation fo the best interests of the principals in good faith.

When the behavior of the system is an AI, the system
can be trained to optimize this proxy objective function.
In the reinforcement learning context, for example,
the objective function can be used as the reward function with which the agent's policy is trained \cite{leike2018scalable}.
If there are hierarchical fiduciary duties between multiple classes of principals, lexicographic \cite{stadler1988fundamentals} or hierarchical \cite{osyczka1984multicriterion} methods in multi-objective optimization can be used to maintain this prioritization.

In the broader research program of AI Alignment \cite{dewey2011learning, hadfield2016cooperative, everitt2021reward}, 
it is understood that alignment through proxy objective functions can fail due to errors.
\citet{leike2017ai} distinguish between the \emph{specification} problem -- when the proxy function is different from the 'true' principal objectives -- and the \emph{robustness} problem -- when the proxy function is accurate, but other problems in the implementation of the system lead to misaligned behavior.
The requirements of specification are uncompromising;
\citet{zhuang2020consequences} demonstrate that incompleteness -- under-specification -- of the proxy objective function can lead to arbitrarily costly behavior of the agent.
Robustness problems arise not from misalignment but from other forms of error in system calibration, such as distributional shift.
In Section \ref{tai:care}, we suggest that mediating these problems be understood as a requirement of the duty of care.

\citet{aguirre2020ai} and others have drawn a connection between
the alignment implied by a duty of loyalty and what is studied more broadly as AGI Alignment.
We maintain that Fiduciary AI is sufficient, but not necessary, for AI Alignment.
Fiduciary AI is a stricter demand than mere alignment because of the constraints from contextual scope.
For example, when \emph{data minimization} -- the requirement that data be collected and stored only when necessary to pursue legitimate purposes -- is a subsidiary duty of loyalty, this limits the power of an AI to act beyond its intended design.
An AGI that is well designed as a Fiduciary AI would, as a matter of its own duties, limit the vulnerability of its principals by performing only within its narrowly defined role \cite{hartzog2022legislating}.\footnote{Perhaps there is one exception. An AI that served as a corporate director, with fiduciary duties to its shareholders, may require general intelligence to operate many aspects of the corporation, even though their duties are narrowly construed in terms of protecting the value of the shareholder's investment in the corporation.
Indeed, this sweeping scope of AI power, combined with the narrowness of its objective function, makes this form of AI particularly risky \cite{russell2019human, benthall2021artificial}.}

\subsection{Subsidiary duties of loyalty}
\label{sec:loyalty-subsidiary}

When a designer schematizes the context of a Fiduciary AI system,
they should enumerate any context-specific norms and rules (Section \ref{tai:context}).
These include subsidiary duties of loyalty, the second part of the ``loyalty two-step'', and can include other TAI standards such as:
data minimization, nondiscriminatory targeting, reducing data sharing with third parties, absence of dark patterns, and content moderation \cite{hartzog2022legislating}. (See Table \ref{tab:subsidiary}.)

Some subsidiary duties may reference the best interests of the principals, but in more nuanced ways than broad ``alignment''.
For example,
the \emph{duty of disclosure} is a positive rule mandating that the fiduciary reveal relevant facts to the beneficiary, especially if these are facts pertaining to a potential conflict of interest.
This duty creates an opportunity for the principal to contest an action that is in potential conflict.
\footnote{\citet{aguirre2020ai} argue that, even in absence of stronger fiduciary duties, for AI the presence of conflicts in design must be ``transparently and saliently indicate[d] to users'', but not forbidden outright. This amounts to extending the duty of disclosure to all AI systems with users.}
Fiduciary AI designers might engage such a duty when, for example,
a system identifies that its objective function is incomplete with respect to its current decision or situation \cite{zhuang2020consequences}, and more information from the principals are needed to guarantee a lack of conflict.
Inversely, the duty of confidentiality is a negative rule prohibiting the disclosure of information (typically, about the beneficiary) for the benefit of the fiduciary.
This subsidiary duty involves representations of both the principal's interests \emph{as well as} the agent's ``true'' interests, such as the interests of its shareholders.

Notably, these subsidiary duties govern flows of information to and from the agent.
They can be understood as transmission principles —
norms governing flows of information as per Contextual Integrity\cite{nissenbaum2020privacy}.
Formal work on the strategic value of information
and its import for AI alignment has been pursued by
\citet{everitt2021agent}.

\section{Care}
\label{tai:care}

Some jurisdictions and contexts define a fiduciary duty of care.
The duty of care obliges the agent to prevent any foreseeable harms to the principal.
It specifically holds the agent liable for accidental harms to a higher degree than they would be held otherwise under tort law.
Whereas loyalty requires the agent to make decisions in the best interest of the principal,
care requires the agent to be \emph{highly informed} before making decisions, so as not to make a decision recklessly.

Fiduciary agents are held to a standard of prudence that is defined contextually.
For example, the executor of a trust is held to the prudent investor rule, which assumes expertise in investing, and is a higher standard than the more general reasonable person standard used in general tort cases.
This is contextually informed level of responsibility 
becomes a field-specific subsidiary duty of care.

AI Safety researchers have identified many potential AI errors that could cause accidental harms\cite{amodei2016concrete}, including:
\begin{itemize}
    \item negative side effects, when an agent disturbs its environment in negative ways not accounted for in its reward function \cite{armstrong2017low, krakovna2018penalizing}
    \item reward hacking, when an agent finds a way to optimize its reward in an unexpected and unintended way  \cite{yuan2019novel, gilbert2022choices}
    \item distributional shift, when the agent is trained in an environment which does not generalize to the environment in which it operates \cite{kirschner2020distributionally, krueger2020hidden}
\end{itemize}
A full discussion of these errors is beyond the scope of this paper.
Here, we note that the fiduciary duty of care is a legal requirement that may include remediation of any or all of these problems as the context-appropriate standard of prudence evolves to take them into account.

For Fiduciary AI, the duty of care may be violated
if the designer's judgment was in bad faith or not
prudently informed of the facts.\footnote{In analogous duty of care cases, 
corporate directors are only held to the standard of making the decision 
with a good faith business judgement, even if their judgement was \emph(ex post) incorrect.
The burden is on the plaintiff to show bad faith or impropriety on the part of the director.
This is called the \emph{business judgement rule}.}
For example, a duty of care for Fiduciary AI might require that the AI designer has considered all potential harms caused by their system.
For example, consider what happens if a hypothetical social media service, \emph{Zmeta}, 
is tried in court for violation of fiduciary duties after it is found to be addictive and harmful to its users.
The company's lawyers argue that engagement is a signal of user interest, and so the company was acting in good faith.
A court might consider a \emph{prudent user rule} informed
by research about addiction and well-being
(such as \cite{tromholt2016facebook, polites2018understanding, kuem2020smartphone}).
Under such a rule, using engagement as a proxy for user
best interest would be considered negligent,
because it falls below the standard of care appropriate
for the context.

Several reporting or labeling frameworks have been proposed for guiding due diligence and accountability for machine learning
based on the data set they use for training \cite{gebru2021datasheets}, conformity with testing standards \cite{arnold2019factsheets}, use of the model only for its intended purposes and with testing for fairness on varying demographic groups \cite{mitchell2019model}, and for overall design of the reward system \cite{gilbert2022choices}.
This can be framed as an improvement to AI Safety due to the reduction of epistemic uncertainty \cite{moller2012concepts, varshney2016engineering, varshney2017safety}, which is uncertainty about which what can be known in principle but has not been discovered in practice.
These standards are likely to evolve over time as AI, and society's understanding of it, matures.

\subsection{Inductive bias}
\label{sec:inductive}

While Fiduciary AI duties apply to everything about a system, from design to deployment, some aspects of it are especially prone to negligence.
Whereas the choice of training data and benchmarking of a model are often deliberate choices, the inductive bias of a machine learning system can be an afterthought, determined by defaults.
This makes attention to inductive bias a good example of what is addressed by the duty of care.

We illustrate this simply in Bayesian terms.
For a particular application of the system
that uses the model trained on a data set $D$ to determine the outcome for a particular
case $h$. For simplicity, let $h$ be a binary decision, such as whether or not to hire an individual. The closer that the likelihood ratio $\frac{P(D | h = 1)}{P(D | h = 0}$ is to 1, the less the available training data has informed the
decision, and the more the inductive bias $P(h)$ dominates.
This can happen when there is insufficient data in the region
that is informative for the case of $h$, such as if the person $h$ is a rare minority among a larger population.
A duty of care for Fiduciary AI would standardize the inductive bias of systems based on their context and purpose of use.
Notably, choices about inductive bias must be made if reward modeling is used to assess the best interests of the principal (see Section \ref{ref:reward-learning}).
\citet{xu2019learning} show how training on other tasks can be used to inform the prior used in IRL on a new task.
Hence, a duty of care might standardize a prior over rewards directly, but it might also standardize a corpus of analogous tasks used to train that prior.

\section{Discussion}
\label{sec:end}

Lawmakers have struggled to find a way to regulate artificial intelligence directly.
At the heart of the problem is the incompleteness with which principals can come to agreement with an AI system, whether through explicit direction or consent \cite{ hadfield2019incomplete}.
This means that for the foreseeable future, AI will only be as ethical as the purposes of the social actor that operates it \cite{benthall2021artificial}.
Fiduciary duties are a time-tested legal means for establishing the trustworthiness of an incompletely contracted agent by aligning its purposes with a principal \cite{easterbrook1993contract}.
Fiduciary duties for computational systems
are part of the law today, and may become more broadly applied with new legislation in the future \cite{balkin2015information, richards2020duty}

This article has aimed to inform system designers about the technical requirements implied by fiduciary duties.
To this end, we have outlined the legal rationale for fiduciary duties, and in particular their application to computational systems and AI.
We have then provided a guide for how a Fiduciary AI system can be designed or audited for compliance with these rules.
This six-step process involves: understanding the context of the system; identifying the principals; assessing the best interests of the principals; aggregating those best interests; ensuring the system is loyal to those best interests; and observing a contextually appropriate standard of care with respect to unlearned aspects of the system, such as inductive bias.

We have identified dimensions of Trustworthy AI that inform best practices in each of these stages. Notably, we see Fiduciary AI practice as drawing on both Contextual Integrity \cite{nissenbaum2020privacy} and AI Alignment \cite{yudkowsky2016ai} research in an original way, while also incorporating other dimensions of Trustworthy AI.
Though Fiduciary AI functions may one day be performed by a general AI system, its duties are scoped to a particular legal context in order to prevent conflicts of interest with the principals and risk more generally.
The legal structure of fiduciary duties -- 
abstract primary duties and more contextually specific subsidiary duties
-- is an important feature of Fiduciary AI, as in many cases legal constraints can make complex AI design challenges more tractable or straightforward.


There are many open research questions raised by the application
of fiduciary principles to computational systems.
We leave as an open question for future work how human-centered
and participatory design methods \cite{martin2020participatory, katell2020toward, birhane2022power, sloane2022participation} might, or might not, be used
to assess the best interests of the principals of an information fiduciary.
We also wonder how the narrowly legal duty of care discussed in
this paper compares with other notions of care in AI \cite{knowles2023trustworthy} and digital infrastructure \cite{tseng2022care} design.
Our focus on AI and legal methods, to the exclusion of more humanistic methods, is a limitation of this article.

Fiduciary AI standards will necessarily evolve with the maturation of policy, technology, and the understanding of the courts.
It is pragmatic to take stock of the state of the art in legal and technical research at this stage so that different research communities can find common ground in Fiduciary AI research, and demonstrate the feasibility of policy positions that anticipate
future iterations of AI research and practice.
Such policies may have far-reaching implications for value alignment of more powerful AI to come  \cite{soares2014aligning, aguirre2020ai}.

\begin{acks}
We gratefully acknowledge the helpful comments of
Andrew Critch,
Francesca Episcopo,
Katrina Geddes,
Jiaying Jiang,
Tomer Kenneth,
Aniket Kesari,
Galen Panger,
Ira Rubenstein,
Katherine Strandburg,
Thomas Streinz,
Emily Tseng,
Kush Varshney,
Mark Verstraete,
Stav Zeitouni.
We also are grateful for the feedback from participants in
the 3rd Annual Symposium for Applications of Contextual Integrity,
Center for Human-Compatible Artificial Intelligence Beneficial AI Seminar,
Northeast Privacy Scholars Workshop,
and the Information Law Institute and Privacy Research Group
at New York University School of Law.
This material is based upon work supported by
the National Science Foundation under Grant No. 2105301. 
Any opinions, findings, and conclusions or recommendations expressed in this paper are those of the authors and do not necessarily reflect the views of the sponsors.
One of the authors of this article was supported by the New York University Information Law Institute’s Fellows program, which is funded in part by Microsoft Corporation.
\end{acks}

\bibliographystyle{ACM-Reference-Format}
\bibliography{fiduciaryai}

\end{document}